\documentclass[12pt]{iopart}
\usepackage{graphicx}
\usepackage{float}
\begin{document}

\title{Pumping vortex into a Bose-Einstein condensate of heteronuclear molecules}

\author{Z F Xu$^1$, R Q Wang$^2$, and L You$^3$}
\address{$^1$Center for Advanced Study, Tsinghua University,\\ Beijing 100084,
People's Republic of China}
\address{$^2$Institute of Physics, Chinese Academy of Sciences,\\ Beijing
100080, People's Republic of China}
\address{$^3$School of Physics, Georgia Institute of Technology, Atlanta,
Georgia 30332, USA}

\ead{zfxu83@gmail.com}

\begin{abstract}Heteronuclear molecules attract wide attention
due to their permanent electric dipole moments. Analogous to atoms
with magnetic dipoles, the existence of nonzero electric dipoles
significantly enhance possibilities and mechanisms for control and
design of quantum degenerate molecule systems with electric (E-)
fields. This work proposes a vortex creation mechanism inside a
condensate of heteronuclear molecules through the adiabatic flipping
of the axial bias of an analogous E-field Ioffe-Pritchard trap
(IPT), extending the original protocol due to Ishoshima, {\it et al.},
[Phys. Rev. A \textbf{61}, 063610 (2000)] for an atomic spinor
condensate inside a magnetic (B-) field IPT. We provide both
analytic proof and numerical simulations to illustrate the high
fidelity operation of this vortex pump protocol. We hope our work
provides stimulating experimental possibilities for the active
investigations in quantum degenerate molecule systems.
\end{abstract}

\maketitle

\section{Introduction}
The impressive progress in atomic quantum gases can be largely
attributed to technical advances in the control of atomic quantum
states utilizing electromagnetic interactions. While laser cooling
with near resonant light narrows down atomic momentum distribution,
forced evaporation on trapped atoms favors spatial confinement into
the quantum degenerate regime. Since the first successful experiment
on atomic Bose-Einstein condensation, or the controlled production
of many bosonic atoms into the same quantum ground state of a
trapping potential, the field of atomic quantum gases has blossomed
into one of the leading frontiers in physics, attracting ever increasing
interest. The ultimate goal along this line of thought
is to control the state of
an arbitrary assembly of an interacting many body system, and either use
it to simulate other physical systems or to perform quantum
computations.

Significant efforts have been directed toward the above ambitious goal,
with the immediate pursuit from more recent endeavors targeted
at the generation, control, and
study of cold molecules. These endeavors include important successes
such as the BCS to BEC cross-over due to tunable interactions
induced through B-field magnetic Feshbach resonances and the
production of quantum degenerate heteronuclear molecules in
specific low lying ro-vibrational states of
their electronic ground states \cite{ni08}. The latter advance
is especially illuminating, highlighting the
prospects for observing and utilizing
anisotropic and long range interactions from the permanent electric dipoles.

Molecule quantum fluids bring in new possibilities due to their
richer internal degrees of freedom, such as alignment and rotation
\cite{friedrich95}, even after their hyperfine spins like those of
atomic spinor quantum gases are neglected. In this short note, we
propose a simple mechanism to generate a vortex state, relying on
the internal degrees of freedom of a heternuclear molecule, through
adiabatically flipping the axial bias of an external E-field
Ioffe-Pritchard trap (IPT) \cite{pritchard83}. The
physics of our proposal can be analogously mapped onto the
successful protocol first proposed \cite{isoshima00} and
demonstrated for magnetically trapped atomic condensates
\cite{leanhardt02,leanhardt03,isoshima07}. Our theory hopefully will
lead to parallel experiments that begin to address new opportunities
afforded by the molecular internal degrees of freedom.

This paper is organized as follows. First, we describe our model for
dc E-field trapping of cold heteronuclear molecules based on the interaction
between their permanent electric dipoles with a spatially varying static
E-field IPT. This is analogous to the operation of the
famous IPT for magnetic dipoles with
an inhomogeneous B-field \cite{pritchard83}. The heteronuclear diatomic molecule is
modeled as a 3D rigid rotor \cite{friedrich95}. We then discuss the eigen structure of
a heteronuclear molecule inside a dc E-field and identify states capable
of trapped by a local E-field minimum and their associated quantum
numbers. Next, the rotational properties of such eigen states are
studied, which helps to illustrate the proposed vortex pump mechanism based
on the flipping of the axial bias E-field
\cite{isoshima00,leanhardt02,leanhardt03,isoshima07,mottonen07,xu08}. Finally we conclude.

It is imperative to present
this study because this analogy between E-field and B-field
and between molecular E-dipole and atomic B-dipole is not a simple one to one map.
The magnetic dipole of an atom contains both an electronic and
a nuclear component, respectively, proportional to the electronic spin ($\vec S$) and
the nuclear spin ($\vec I$). Its projection onto the local B-field is
approximately proportional to the projection of its hyperfine spin $\vec F=\vec S+\vec I$
when the B-field is weak. In contrast, the
molecular dipole is a fixed constant and pointed along the molecule
axis connecting the two nuclei. The internal
rotational angular momentum of the molecule stays in the perpendicular
plane of the molecule axis in the absence of an E-field. Despite
these differences, our study finds that the Isoshima, {\it et al.},
protocol \cite{isoshima00,leanhardt02,leanhardt03},
 originally developed for B-field trapped atoms or
molecules, remains effective for heteronuclear molecules in a
E-field IPT. Much of these comparative studies and analogies will
become clear from the materials presented in the third and fourth
sections.

\section{Trapping a heteronuclear molecule with static electric field}

We consider a diatomic molecule composed of two different atomic
nuclei and described in terms of their center of mass $\vec R$ and
relative coordinate $\vec r$. Neglecting the nuclear and electronic
spins and their associated interactions, or assume we consider an eigen
state of the above internal degrees of freedom, the remaining
Hamiltonian reads
\begin{eqnarray}
  \mathcal{H}&=&\frac{\mathbf{P}_{R}^2}{2M}+\frac{\mathbf{P}^2_{r}}{2\mu}
  -\vec{D}(\vec r)\cdot\vec{E}(\vec{R})
  \nonumber\\
  &=&\frac{\mathbf{P}_{R}^2}{2M}-\frac{\hbar^2}{2\mu r^2}\frac{\partial}{\partial r}
  \left( r^2\frac{\partial}{\partial r} \right)+\frac{\mathbf{J}^2(\theta_r,\phi_r)}{2\mu r^2}
  -\vec{D}(\vec r)\cdot\vec{E}(\vec{R}),
  \label{ham}
\end{eqnarray}
where $\vec{E}(\vec{R})$ denotes the inhomogeneous static E-field,
which is slowly varying over the molecule size of $r$, and
$\vec{D }(\vec r)$ is the permanent electric dipole moment operator
for this specific electronic and internal spin eigen-state \cite{friedrich95,cohen84,micheli07}.
$M=M_1+M_2$ and $\mu=M_1M_2/(M_1+M_2)$ are, respectively, the total
and the reduced mass of the two atoms making up the molecule.

We adopt the Bohn-Oppenheimer approximation to study the effective
trapping of the molecule center of mass motion. Within this
approximation, the molecular wave function for $\vec R$ and $\vec r$
is decomposed into the form
$\Psi(\vec{r},\vec{R})=\Phi(\vec{r})\psi(\vec{R})$, and we find
\begin{eqnarray}
  \fl\qquad\left[-\frac{\hbar^2}{2\mu r^2}\frac{\partial}{\partial r}
  \left( r^2\frac{\partial}{\partial r} \right)+\frac{\mathbf{J}^2(\theta_r,\phi_r)}{2\mu r^2}
  -\vec{D}(\vec r)\cdot\vec{E}(\vec{R})\right]\psi_{M_J}(\vec{r})
  &=&\mathcal{V}_{M_J}(\vec{R})\psi_{M_J}(\vec{r}), \label{rot}\\
  {\hskip 112pt \left(\frac{\mathbf{P}_{R}^2}{2M}+\mathcal{V}_{M_J}(\vec{R})\right)
  \Phi_n(\vec{R})}&=&E_{M_J}^{(n)}\Phi_n(\vec{R}).
\end{eqnarray}

Now consider for a specific vibrational state, the Eq. (\ref{rot})
above then reduces to the simple form of a 3D rotator in a dc E-field
with a rotational constant $B=\langle\hbar^2/2\mu r^2\rangle$. If we
further assume the local E-field direction to be the quantization
z-axis, then the Schr\"odinger equation for the internal part of the
diatomic molecule becomes
\begin{eqnarray}
  \left[BJ^2-DE(\vec R)\cos\theta_r\right]\psi_{M_J}(\theta_r,\phi_r)
  =\mathcal{V}_{M_J}(\vec R)\psi_{M_J}(\theta_r,\phi_r),
\end{eqnarray}
where we have neglected the dependence of the permanent dipole moment on the
internuclear distance $\vec r$.

Some intuition can be gained for weak E-fields if a simple perturbation
theory is adopted as is done in Ref. \cite{micheli07}. We take the
unperturbed Hamiltonian to be $H_0=BJ^2$ with
$H_0|J,M_J\rangle=E_{JM_J}|J,M_J\rangle$, where the eigen-energy and
eigen-function are, respectively, $E_{JM_J}=BJ(J+1)$
and $|J,M_J\rangle$, with spherical harmonics $|J,M_J\rangle$ being
functions of $\hat r$.
The dipole interaction $-DE(\vec R)\cos\theta_r$
is treated as a perturbation.
To first order the wave function becomes
\begin{eqnarray}
  \psi_{M_J}&=&|J,M_J\rangle+\frac{DE}{B}\frac{1}{2(J+1)}
  \sqrt{\frac{(J+1)^2-M_J^2}{(2J+1)(2J+3)}}\,|J+1,M_J\rangle \nonumber\\
   && -\frac{DE}{B}\frac{1}{2J}\sqrt{\frac{J^2-M_J^2}{(2J-1)(2J+1)}}\,|J-1,M_J\rangle,
   \label{psifo}
\end{eqnarray}
and the corresponding eigen-energy to second order in the external E-field
becomes
\begin{eqnarray}
  \mathcal{V}_{M_J}&=&BJ(J+1)+\frac{D^2E^2}{B}\frac{1-3M_J^2/J(J+1)}{2(2J-1)(2J+3)}.
\end{eqnarray}

For this model system, $M_J$ is a conserved quantity.
Assuming the center of mass motion is adiabatic with respect to the
internal state, the above eigen energy from the internal state
$\mathcal{V}_{M_J}$ then clearly acts as an effective potential
due to its dependence on $\vec R$. In particular, we find any state with
$1-3M_J^2/J(J+1)>0$ is a weak field seeking state, {\it i.e.}, free
space E-field traps analogous to the B-field IPT can be constructed.
These states, of course, are meta-stable, and trapping is
possible because of the dynamic stability from the Larmor-like
precession of the permanent electric dipole along the direction of
a local E-field. On the other hand, the strong field seeking states with
$1-3M_J^2/J(J+1)<0$ cannot be used to spatially confine
molecules with static fields, because it is impossible to construct a local
E-field maximum.

The significant progress gained over the years
in trapping and manipulating neutral atoms with static B-fields \cite{TB}
provide important enabling technologies for the experimental successes of
cold atom physics research. A variety of B-field traps have been implemented
for various applications \cite{pritchard83,TB}.
Despite the analogy described earlier
of the B-field confinement of magnetic dipoles with
the E-field confinement of electric dipoles (of polar molecules),
experimental efforts of trapping and manipulating polar molecules
with electrostatic E-fields become an active research topic
only in recent years with the interests on cold atoms broadening into
cold molecules. The Meijer's group pioneered the research of cooling and
 trapping of molecules with (3D quadrupole) E-fields \cite{GM00}.
They also demonstrated the guiding of
cold molecules with a torus shaped (2D) E-field hexapole \cite{GM01}.
With suitable modifications, an analogous IPT of E-field can be
constructed for polar molecules \cite{pritchard83,TB,ab,GM}.

\begin{figure}[H]
\centering
\includegraphics[width=4.5in]{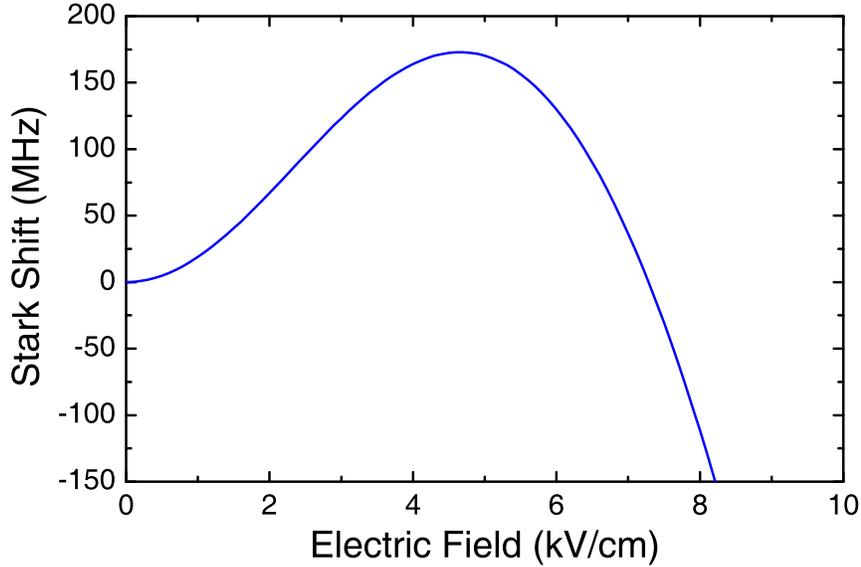}
\caption{The Stark shift of the $M_J=1$ state for the polar molecule KRb in the
electronic ground state $X^1\sum{\nu=0}$, for realistic electrostatic
E-trap field strength \cite{GM,ab}. }
\label{fig1}
\end{figure}

In Fig. \ref{fig1}, we show the calculated dc Stark shift for
the $M_J=1$ state of heteronuclear molecule KRb
that is connected to the $J=2$ manifold when the E-field is absent.
We adopt the
parameters as from its ground state $X^1\sum{\nu=0}$ as
measured recently \cite{ni08} with the electric dipole
moment $D=1.36\times 10^4$\,$\rm Hz\cdot m/V$ and the
rotational constant $B=1.1139\, \rm GHz$.
The calculation is accomplished by a numerical diagonalization of the
Hamiltonian over a truncated basis of common eigen states for
operators $\mathbf{J}^2$ and $J_z$. The diagonalization
is carried out in the subspace of a conserved
$M_J$ \cite{cohen84}, and the procedure is found to be rapidly
convergent. At the E-field strength we consider, the
perturbation result $\psi_{M_J}$ in Eq. (\ref{psifo}) with $M_J=1$ and $J=2$
is found to consists of an excellent approximation. This particular state
as illustrated in Fig. \ref{fig1} is clearly a weak field seeking state.
The maximum trap height can become as high as $173\,\rm MHz$.

\section{Rotational transformation of the wave function $|J,M_J\rangle$}
Before we extend the Isoshima, {\it et al.}, protocol \cite{isoshima00}
from B-field trapped atoms to E-field trapped polar
molecules, we will briefly review the rotational properties of the
spherical harmonics $|J,M_J\rangle$. Under a rotation
along a unit vector $\hat{n}$ by an angle $\theta$,
the wave function changes to
\begin{eqnarray}
  R(\theta\hat{n})|J,M\rangle=e^{-i\theta\hat{n}\cdot\mathbf{J}}|J,M\rangle.
\end{eqnarray}
Expressed in terms of the Euler angles, the most general 3D rotation
takes the following form
\begin{eqnarray}
  R(\alpha,\beta,\gamma)=\exp(-i\alpha J_z)\exp(-i\beta J_y)\exp(-i\gamma J_z).
\end{eqnarray}
According to the representation theory of rotation, we find
\begin{eqnarray}
  R(\alpha,\beta,\gamma)|J,M\rangle&=&\sum\limits_{M'}D^{J}_{M'M}(\alpha,\beta,\gamma)
  |J,M'\rangle\nonumber\\
  &=&\sum\limits_{M'}\exp[-iM'\alpha]d^{J}_{M'M}(\beta)\exp[-iM\gamma]|J,M'\rangle,
\end{eqnarray}
where
\begin{eqnarray}
  d^{J}_{M'M}(\beta)&=&\langle J,M'|\exp[-i\beta J_y]|J,M\rangle\nonumber\\
  &=&[(J+M)!(J-M)!(J+M')!(J-M')!]^{1/2}\nonumber\\
  &&\times\sum\limits_{\nu}[(-1)^{\nu}(J-M'-\nu)!(J+M-\nu)! (\nu+M'-M)!\nu!]^{-1}\nonumber\\
  &&\times\left(\cos\frac{\beta}{2}\right)^{2J+M-M'-2\nu}\left(-\sin\frac{\beta}{2}\right)^{M'-M+2\nu}.
\end{eqnarray}
The value of $\nu$ needs to ensure that the numbers in the factorials
stay non-negative. For example, when $\beta=\pi$ and
$\cos\frac{\beta}{2}=0$, for $d^{J}_{M'M}$ to be nonzero, we need to
let $2J+M-M'-2\nu=0$. This gives
\begin{eqnarray}
  (J-M'+\nu)!=\left[ -\frac{1}{2}(M'+M) \right]!,\\
  (J+M-\nu)!=\left[ \frac{1}{2}(M'+M) \right]!.
\end{eqnarray}
In order to assure that both $J-M'-\nu$ and $J+M-\nu$ are
non-negative, we need to make $M'+M=0$. As a result we find
\begin{eqnarray}
  d^{J}_{M'M}(\pi)=(-1)^{J-M'}\delta_{M',-M},
\end{eqnarray}
which gives
\begin{eqnarray}
  R(\alpha,\pi,\gamma)|J,M\rangle=(-1)^{J+M}\exp[iM\alpha]\exp[-iM\gamma]|J,-M\rangle.
\end{eqnarray}

We want to stress at this point that the angles $\alpha$, $\beta$,
and $\gamma$ are parameters used to specify arbitrary rotations, and
they have nothing to do with the internal state angles $\theta_r$
and $\phi_r$. To understand the physics behind the vortex generation protocol,
which is directly related to the geometrical properties of the
static E-field, we first consider several special cases of
unit vectors as rotation axes: for
$\hat{n}=\hat{X}\cos\phi+\hat{Y}\sin\phi$,
$\hat{n}_{\perp}=-\hat{X}\sin\phi+\hat{Y}\cos\phi$, and
$\hat{n}_q=\hat{X}\sin\phi+\hat{Y}\cos\phi$, respectively, with $\hat{X}$ and
$\hat{Y}$ the unit vectors along the Cartesian $x$- and
$y$-directions. $\theta$ and $\phi$ denote the polar and
azimuthal angles of the center of mass coordinate $\vec R$. The last
case of $\hat{n}_q$ corresponds simply to the direction of a IPT
E-field in the plane of $Z=0$. Using the
relationship derived above, we find that
\begin{eqnarray}
  R_{\hat{n}}(\pi)|J,M_J\rangle&=&\exp(-i\pi\hat{n}\cdot \mathbf{J})|J,M_J\rangle\nonumber\\
  &=&R_{\hat{z}}(\phi-\pi/2)R_{\hat{y}}(\pi)R_{\hat{z}}(\pi/2-\phi)|J,M_J\rangle\nonumber\\
  &=&-(-1)^{J+M_J}\exp[2iM_J\phi]|J,-M_J\rangle, \\
    R_{\hat{n}_{\perp}}(\pi)|J,M_J\rangle&=&\exp(-i\pi\hat{n}_{\perp}\cdot \mathbf{J})|J,M_J\rangle\nonumber\\
  &=&R_{\hat{z}}( (\pi/2+\phi)-\pi/2)R_{\hat{y}}(\pi)R_{\hat{z}}(\pi/2-(\pi/2+\phi))|J,M_J\rangle\nonumber\\
  &=&(-1)^{J+M_J}\exp[2iM_J\phi]|J,-M_J\rangle,\\
  R_{\hat{n}_q}(\pi)|J,M_J\rangle&=&\exp(-i\pi\hat{n}_q\cdot \mathbf{J})|J,M_J\rangle\nonumber\\
  &=&R_{\hat{z}}( (\pi/2-\phi)-\pi/2)R_{\hat{y}}(\pi)R_{\hat{z}}(\pi/2-(\pi/2-\phi))|J,M_J\rangle\nonumber\\
  &=&(-1)^{J+M_J}\exp[-2iM_J\phi]|J,-M_J\rangle.\label{rotipt}
\end{eqnarray}

These properties of rotational transformation show that by
enforcing a rotation of the internal state, the wave function gains
an appropriate topologically phase specified by the
azimuthal angle coordinate of the center of mass (or the molecule) coordinate.
Provided the internal state flipping of
the permanent dipole moment and the center of mass motion are both
adiabatic, the above results show that different vortical phases are generated
as in the protocol of Isoshima, {\it et al.}, for B-field trapped
atomic spinor condensates \cite{isoshima00,leanhardt02,leanhardt03}.

\section{Vortex creation in a condensate of heteronuclear molecules}

In this section, we will confirm numerically the vortex creation
protocol for a condensate of heteronuclear molecules in an E-field
IPT. Basically, we will simulate the axial E-field bias flip and
check for the adiabatic conditions of the associated internal state.
There are two main points we need to watch for. First we need to
create the proper vortical phase structure on a condensate during
the time evolution of the E-field. Secondly we must maintain
adiabaticity during the flip of the internal state of a polar
molecule.

Considering the first question, assume the initial state
corresponds to a polar molecule locally aligned along the E-field
direction of a E-field IPT, which is essentially along the axial $z$-axis direction
close to the $z$-axis, its local internal state $\phi_{M_J}(\theta_r,\phi_r)$
can then be described approximately as the
eigen state of the system in the local E-field. Since
$M_J$ is a conserved quantity, this eigen state $\psi_{M_J}(\theta_r,\phi_r)$
can be expanded by a series of $|J,M_J\rangle$ in $J$,
\begin{eqnarray}
  \psi_{M_J}(\theta_r,\phi_r)=\sum\limits_J C_J|J,M_J\rangle.
  \label{psimj}
\end{eqnarray}
The flipping of the E-field bias adiabatically corresponds then
to nothing but a rotation of the initial state along
the unit vector $\hat{n}_q(\theta,\phi)$ in the transverse plane
by an angle of $\pi$. According to the rotation properties
we discussed before, this flipping of the bias then gives
\begin{eqnarray}
  R_{\hat{n}_q}(\pi)\psi_{M_J}(\theta_r,\phi_r)=\sum\limits_J (-1)^{J+M_J}
  C_J\exp[-2iM_J\phi]|J,-M_J\rangle,
  \label{rotpsimj}
\end{eqnarray}
which gains clearly a vortex with a winding number $-2M_J$.

Turning to the second question concerning the adiabaticity, we simply can
propagate the initial wave function $\psi_{M_J}$ in real time,
simulating the complete process of the bias flip. To test the level
of adiabaticity and the validity of Eq. (\ref{rotpsimj}), we
consider points $(X_0, Y_0, 0)$ along a circle in the $X$-$Y$ plane.
The absolute value for the E-field along the circle is then a
constant, although their local directions are all different. In
fact, for the E-field IPT being considered here, the E-field can be
written as $\vec{E}=E'(X,-Y,L)$ with $L$ a constant. Choosing the
weak field seeking state $\psi_{M_J=1}$ with $J=2$ at $\vec{E}=3\,
\rm kV/cm\ \hat{Z}$ as considered in Fig. \ref{fig1}, we simulate
the flip protocol over various time intervals. The time evolution of
the bias E-field is taken to be the same as that in the optimal
B-field protocol \cite{xu08}, and the E-field in the transverse
$X$-$Y$ plane is treated as a constant.

\begin{figure}[H]
\centering
\includegraphics[width=4.5in]{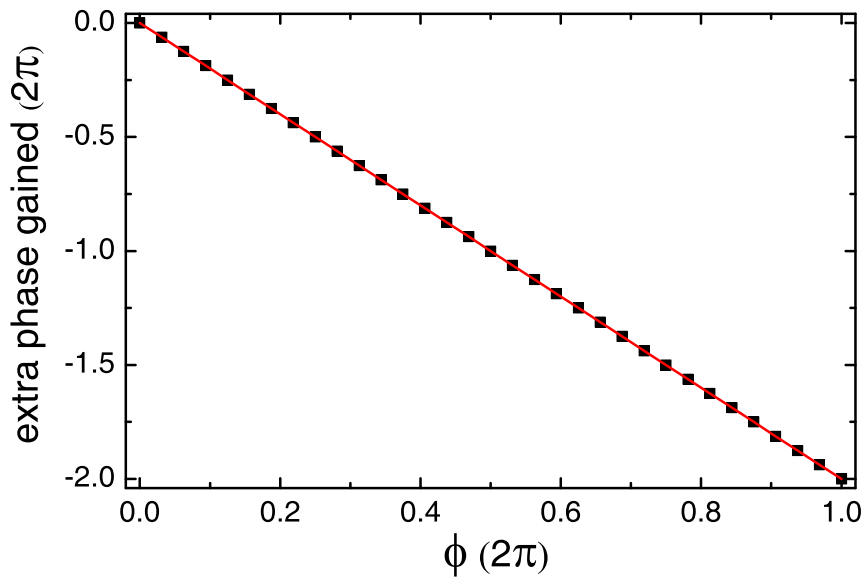}
\caption{The topological phase gained after numerically flipping the
bias E-field, that is calculated by taking the inner product of the
time evolved state after flip with the nominally down polarized
($M_J=-1$) state. $\fullsquare$ denotes numerically computed data,
and the red solid curve represents a linear fit. Both axes are
dimensionless in units of $2\pi$.  }
\label{fig2}
\end{figure}

In our simulation, we take the E-field at $t=0$ as $E'L=3\, \rm
kV/cm$ and $E'\sqrt{X_0^2+Y_0^2}=0.3\, \rm kV/cm$. After the flip of
the E-bias over $10^5/B$ ($\sim 0.1$ ms at $B=1.1139\, \rm GHz$), which is
a reasonably
fast time scale, we calculate the inner product of the final state
with the eigen-state $\psi_{M_J=-1}$ (of $J=2$) when the E-field
$\vec{E}=-3 \rm\,kV/cm\,\hat{Z}$ is pointed in the downward
direction. Using these parameters, the adiabaticity is found to be
fully satisfied, and the state overlap is essentially 100\% except
for the extra topological phase gained as shown in Fig. \ref{fig2}.
A numerical fit gives precisely the slope for the phase over $\phi$
being exactly equal to $-2$, which confirms the high fidelity
operation of our vortex pump proposal.

Before concluding, we point out that, like the system of
magnetically trapped atomic spinors inside a B-field IPT
\cite{peng07,xu08}, the quantity of $J_z-L_z$ is found to commute
with the Hamiltonian $B\mathbf{J}^2-\mathbf{D}\cdot E'(X,-Y,L)$,
where $E'$ is the spatial gradient of the E-field IPT, and $L_z$ is
the mechanical angular momentum of the heteronuclear molecule. $\vec
J$ is the rotational angular momentum of the molecule. If the flip
of the E-bias is indeed adiabatic, the respective quantum numbers
conserve the combination $M_J-L_z$. After going through the flip,
the internal rotational state is changed from $M_J$ to $-M_J$, which
then must be accompanied by an increase of $2M_J$ to its mechanical
angular momentum. Thus we see the appearance of a vortex state.

In conclusion, by analogy with B-field trapping of neutral atoms
with magnetic dipoles, we study and identify weak field trapping
states of polar molecules inside a spatially inhomogeneous dc
E-field. Further, we suggest that an effective and efficient vortex
pump protocol can be envisioned for condensates of polar molecules
\cite{mottonen07,xu08}, based on the flipping of the axial bias
field, originally suggested \cite{isoshima00} and experimentally
demonstrated for atomic spinor condensates inside a B-field IPT
\cite{leanhardt02,leanhardt03,isoshima07}. We have confirmed that
the E-field bias flip remains effective, and the vortex state created
maintains a vorticity proportional to the $M_J$ quantum number of
the trapped molecule state. When a diatomic molecule is placed
inside a homogeneous E-field, $M_J$ is a good quantum number. The
weak field trapping states can be associated with very large values
of $M_J$, thus the amount of vorticity created could become very
significant even after a single bias flip. This vorticity possibly could open
up a practical approach to reach the rapid rotation limit of
atomic/molecular quantum gases.

More generally, we find that this protocol for vortex creation
remains effective if the diatomic molecule is taken as a symmetric
top \cite{herzberg}. Interestingly we find the E-bias flip also
works for the strong field trapping states provided the polar
molecules are confined through other means not relying on its
permanent electric dipole. Finally, other improvements to the
 B-field bias flip protocol \cite{isoshima00}, such as
those developed for cyclically operated continuous vortex pumping schemes can
be analogously extended to the case of heteronuclear molecules
\cite{mottonen07,xu08}.

\section{Acknowledgement}
This work is supported by US NSF, NSF of China under Grant
10640420151, and NKBRSF of China under Grants 2006CB921206 and
2006AA06Z104.

\end{document}